\documentclass[aps,pre,11pt,superscriptaddress,floatfix,tightenlines,showpacs,longbibliography,notitlepage]{revtex4-1}
\usepackage{graphicx}
\usepackage{bm}
\usepackage{amsfonts}
\usepackage{color}
\usepackage{amsmath}    
\usepackage{epsfig}
\AtBeginDocument{%
    \newwrite\bibnotes
    \def\bibnotesext{Notes.bib}
    \immediate\openout\bibnotes=\jobname\bibnotesext
    \immediate\write\bibnotes{@CONTROL{REVTEX41Control}}
    \immediate\write\bibnotes{@CONTROL{%
    apsrev41Control,author="08",editor="1",pages="1",title="0",year="1"}}
     \if@filesw
     \immediate\write\@auxout{\string\citation{apsrev41Control}}%
    \fi
}%
\renewcommand{\Re}{\mbox{\textit{Re}}}
\newcommand{\St}{\mbox{\textit{St}}}

\newcommand{\ictsaddress}{International Centre for
  Theoretical Sciences, Tata Institute of Fundamental Research,
  Bangalore 560089, India}
\newcommand{\iiseraddress}{Indian Institute for Science Education and Research, Pune 411008, India}

\begin{document}
\title{Flow structures govern particle collisions in turbulence}
\author{Jason R. Picardo}
\email{jrpicardo@icts.res.in; picardo21@gmail.com}
\affiliation{\ictsaddress}
\author{Lokahith Agasthya}
\email{lokahith.agasthya@students.iiserpune.ac.in}
\affiliation{\iiseraddress}
\author{Rama Govindarajan}
\email{rama@icts.res.in}
\affiliation{\ictsaddress}
\author{Samriddhi Sankar Ray}
\email{samriddhisankarray@gmail.com}
\affiliation{\ictsaddress}
\begin{abstract}

The role of the spatial structure of a turbulent flow in enhancing particle
collision rates in suspensions is an open question. We show and quantify, as a
function of particle inertia, the correlation between the multiscale structures
of turbulence and particle collisions: Straining zones contribute predominantly
to rapid head-on collisions compared to vortical regions. We also discover the
importance of vortex-strain worm-rolls, which goes beyond ideas of preferential
concentration and may explain the rapid growth of aggregates in natural
processes, such as the initiation of rain in warm clouds.

\end{abstract}

\maketitle

\begin{figure}
\includegraphics[width=.55\textwidth]{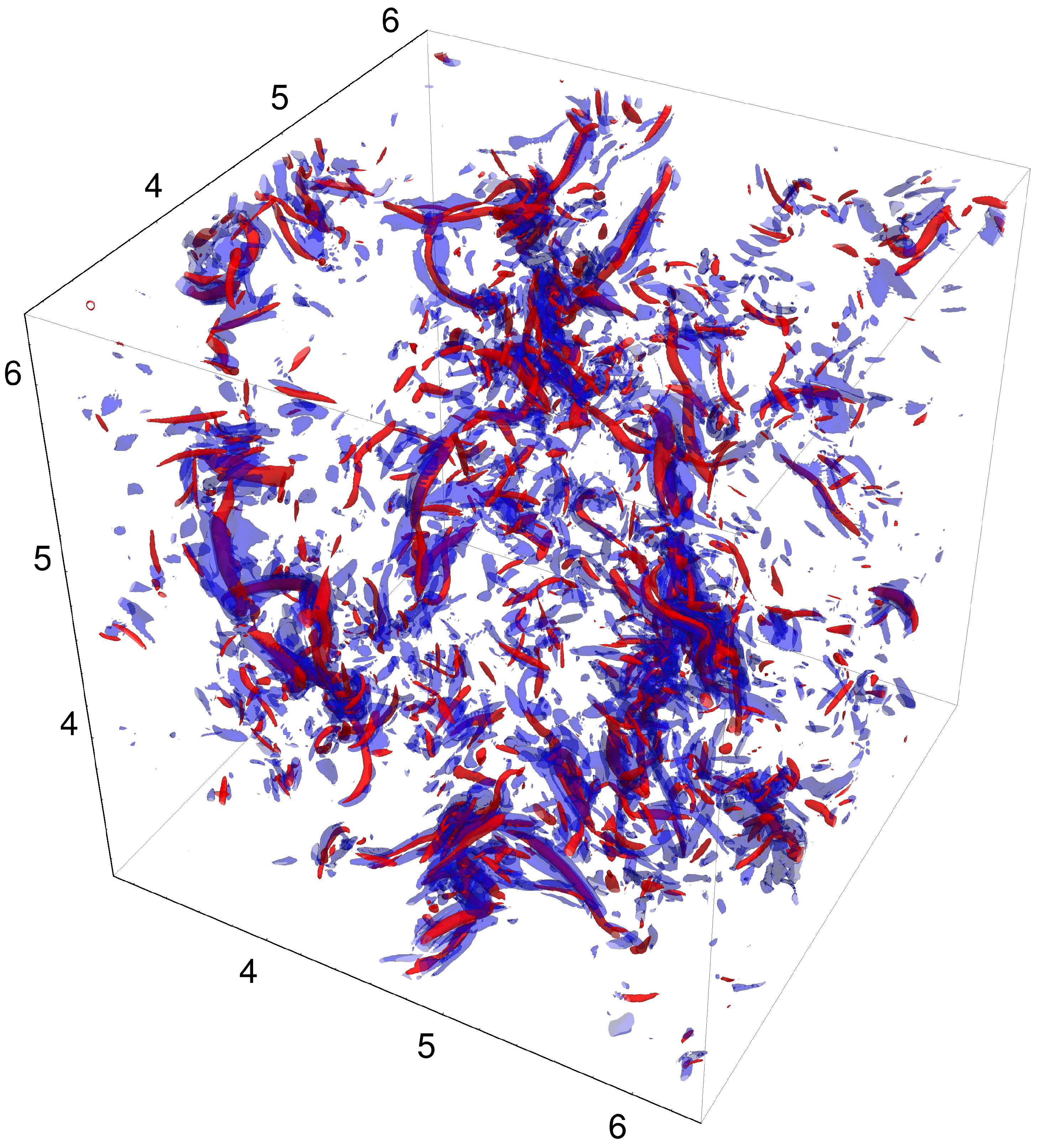}
	\caption{A representative snapshot of three-dimensional contours of ${\mathcal Q}$ showing 
	intense vortex tubes (opaque red: $+5.6 \sqrt{\langle {\mathcal Q^2}\rangle}$)
enveloped by dissipative, straining sheets (transparent blue: $-2 \sqrt{\langle {\mathcal Q^2}\rangle}$).}
\label{fig:Qstruc}
\end{figure}

Turbulence is riddled with a hierarchy of interacting vortical and straining
structures (Fig.~\ref{fig:Qstruc}), which are closely related to its
characteristic intermittent and non-Gaussian statistics
\citep{FrischCUP,Tsinober2009,Schumacher2012,Frisch2012,Buzzicotti2016,Lanotte2016}. 
The most intense structures typically occur near each
other, in the form of vortex tubes surrounded by straining sheets,
as shown in Fig.~\ref{fig:Qstruc}. This organization---a sort of  vortex-strain \textit{worm-rolls}---is characteristic of
turbulent flows~\citep{Zeff2003,Ray2009,Schumacher2010,Gotzfried2017}, and its origin
and dynamical implications continue to be investigated
\citep{Guala2007,Hamlington2008,Ray2015,Wilczek2015,Lawson2015,Ray2018}.
These structures distinguish fully developed turbulence from purely random flow fields,
and must play an important role in many aspects of turbulent transport. 
The most important of these---because it remains central to our understanding of phenomena as 
diverse as the formation of planets in circumstellar disks
\citep{Lissauer2013} or the initiation of rain in warm clouds
\citep{Grabowski2013, Chen2018}---is the growth of macroscopic aggregates, due to collisions and
coalescences, from nuclei-particles (dust or aerosols) suspended in a turbulent flow. 
The role of the underlying turbulent carrier flow is critical: Estimates of, e.g., the rate of 
growth of these aggregates in the absence of such flows do not agree with that seen in nature~\citep{Pumir2016}.
Indeed, the explanation of such rapid growth through coalescence,  demonstrated
\citep{sundaram1997,Bec2016,Onishi2015} and quantified in terms of flow
statistics \citep{vobkuhle2014,Pumir2016,Ireland2016}, is rooted in
the ability of turbulent flows to enhance the rate of collisions between
nuclei seed-particles. 

A critical discovery, due to Bec, \textit{et
al.}~\citep{Bec2016}, was to find the precise connection between the
intermittent (multiscaling) nature of the carrier turbulent flow and the
accelerated growth of aggregates. And yet the implied correlation between the
structure of the flow and droplet collisions-coalescences remains unknown.
Indeed, there is evidence to suggest that flow structures matter. But the
question is how and when?
 
In this paper we answer this question comprehensively, based on 
direct numerical simulations (DNSs), and show how
straining regions are intrinsically more effective at generating collisions
than vortical ones even for uniformly distributed \textit{inertia-less}
particles.  Particle inertia widens this discrepancy, not simply by
preferential concentration, but also by selectively increasing the collision
velocities in straining zones. This is because straining regions have a larger
proportion of \textit{head-on} or \textit{rear-end} collisions, as opposed to
\textit{side-on} collisions, which are predominant in vortical regions.
Consequently, a larger fraction of the velocity gradient in straining zones is
translated into the particle approach velocity.  Finally, and most strikingly,
we show how intense vorticity and strain, cohabiting as vortex-strain \textit{worm-rolls}, conspire to generate rapid,
violent collisions. 


We therefore consider an incompressible ($\nabla\cdot {\bm u} = 0$) turbulent flow whose velocity $\bm u$ 
is a solution to the Navier-Stokes equation
\begin{equation}
\partial_t {\bm u} + ({\bm u} \cdot \nabla){\bm u} = -\nabla p + \nu \nabla^2 {\bm u} + {\bm f}
\end{equation}
where $\nu$ is the kinematic viscosity.
We perform DNSs on a tri-periodic domain with
$N{^3}=512^3$ grid points, by using a standard de-aliased pseudo-spectral solver~\citep{Canuto1_2006} and 
a second-order Adams-Bashford scheme for time-integration. A statistically stationary, homogeneous and isotropic
flow is maintained by the time-dependent large-scale forcing $\bm f$, which
injects a constant amount of energy, and hence dissipation $\epsilon$, into the first two wavenumber shells. 
The Kolmogorov length $\eta = (\nu^3/\epsilon)^{1/4}$ satisfies $\eta k_{\rm
max} \approx 1.7$ (where $k_{\rm max} = \sqrt{2}N/3$ is the maximum resolved
wavenumber). The Taylor-Reynolds number $\Re_\lambda = 2E\sqrt{5/(3 \nu
\epsilon)}$ = 196, where $E$ is the total kinetic energy. The compensated energy spectrum 
from such a simulation shows an inertial range which is certainly less than a decade but still resolved (see, e.g., 
Fig. 1 (a) for $N$ = 512 and $D = 3.00$ in Ref.~\cite{Buzzicotti2016}).

To identify vortical and straining structures in the flow, we use the second invariant of the velocity gradient
tensor ${\mathcal Q} = (R^2 - S^2)/2$~\citep{Dubief2000,Blackburn1996,Chong1990}, defined through the 
velocity gradient tensor ${\mathcal{A}}  = \tau_{\eta}
\nabla {\bm u}$ (normalized by the Kolmogorov time $\tau_\eta=(\nu/\epsilon)^{1/2}$) which yields the
symmetric strain rate tensor ${\mathcal S} = ({\mathcal{A}}
+{\mathcal{A}}^{\mathrm T})/2$ and the anti-symmetric rotation rate tensor
${\mathcal R} = ({\mathcal{A}} -{\mathcal{A}}^{\mathrm T})/2$. Regions where ${\mathcal Q} > 0$ (${\mathcal Q} < 0$) are
dominated by vorticity (irrotational strain)~\citep{Ireland2016}.
Figure \ref{fig:Qstruc} presents contours of large
positive and negative values of ${\mathcal Q}$, which reveal characteristic
vortex-strain worm-rolls. 

We introduce in the flow $10^6$ identical particles, each having a
sub-Kolmogorov radius $a = \eta/3 $  and a density $\rho_p$ much larger than
that of the carrier-fluid $\rho_f$. The particles occupy a small volume
fraction of ${\mathcal O} (10^{-4})$ and their influence on the flow is
negligible. Since the Reynolds number associated with their slip
velocities is small, and $\rho_p \gg \rho_f$, the evolution of particle trajectories $\bm{X}_{p}(t)$
is determined by the simplified Maxey-Riley equations~\citep{CroorRev2017,Armenio2001,Hinsberg2017,Aartrijk2010}:
\begin{equation}
\frac{d \bm{X}_{p}}{dt}=\bm{V}_{p}, \quad\quad \frac{d \bm{V}_{p}}{dt}=-\frac{1}{\tau_{p}} [\bm{V}_{p}-{\bm u}(\bm{X}_{p},t) ] \label{eq:maxey}
\end{equation}
where $\tau_p = 2 a^2 \rho_{p} /(9 \nu \rho_{f})$, the particle relaxation
time, yields the Stokes number ($\St =\tau_p /
\tau_\eta)$, which provides a non-dimensional measure of the particle's
inertia.  We consider several families of particles, with $\St$ ranging from 0
to 16.75 and use an exponential integration scheme \citep{Ireland2013}. The case of tracers $\St =
0$ is handled separately by using a second order Runge-Kutta time-stepper.
The fluid velocity at the particle position is obtained via fourth-order
B-spline interpolation \citep{Hinsberg2012}.

It is important to note that the particle radius influences
particle transport in two distinct ways: It sets the value of $St$, while also
determining the separation between particle centers at collision. As we are considering heavy particles ($\rho_p/\rho_f > 1$), the radius must be very small for small values of $St$. This makes the gathering of collision statistics in this dilute suspension increasingly impractical as $St \rightarrow 0$, unless we consider an artificially enlarged collision radius. Indeed, for tracers, by definition, any choice of radius is arbitrary. Our choice of $\eta/3$ is physically consistent for particles with $St>0.1$. However, for smaller $St$, this particle radius would imply $\rho_p/\rho_f < 1$ which contradicts our heavy particle assumption. Indeed, to accurately describe particle motion, the simplified Maxey-Riley equations \eqref{eq:maxey} require $\rho_p/\rho_f > O(10^2)$ \citep{Armenio2001,Hinsberg2017,Aartrijk2010}. Nevertheless, it is convenient to consider this enlarged collision radius, solely for the purpose of detecting collisions. As a check, we actually
performed simulations with particle radius $\eta/10$ (where the problem is mitigated) and found our results unchanged. This can be rationalised  by noting
that particles move ballistically in the small distance which separates the
$\eta/3$ and the more realistic $\eta/10$ radii and hence make our collision results
insensitive to the precise choice of the radius. Nevertheless, we do report results
for the larger radius because the collision statistics
are much poorer with the smaller radius.  

After the randomly-seeded particles have settled into a statistically
stationary distribution, we begin detecting collisions using an algorithm
similar to that in Ref.~\citep{sundaram1997}.
After a collision is detected, the two particles are allowed to move past each other without any modification to their trajectories. This \textit{ghost collision} approach is a standard approximation that, by ignoring coalescence, allows one to measure collision rates while the particle distribution remains in a statistically stationary state. This feature allows us to identify and compare the regions of the flow where most of the collisions occur to those regions where most of the particles reside. Of course, because particles never coalesce, the collision rates are over-predicted (this issue is less severe for dilute systems, such as the one we consider here). We do not expect this to impact our conclusions, however, as our study is based on identifying where collisions occur, which depends on the relative values of the collision rate in different regions of the flow and not on the absolute values.

The rate of collisions depends, of course, on the relative velocity
of particles at contact~\citep{Saw2014,James17,Bhatnagar2018,Bhatnagar2018b}. For tracers ($\St = 0$) or nearly-tracer particles ($\St \gtrsim 0$),
this is determined by fluid velocity gradients ($\propto \tau_\eta^{-1}$),
which increase in magnitude as the flow becomes more turbulent. This picture, underlying the work of
Saffman and Turner~\citep{saffman1956}, is blind 
to flow structures: It disregards whether the local velocity gradient arises from rotation or strain.

Inertial particles ($\St > 0$), e.g., droplets in air,  
\textit{preferentially concentrate}, thereby increasing their local
number density~\citep{sundaram1997}. They can also attain relative velocities
much larger than that of the underlying flow. Dubbed the \textit{sling effect}
\citep{Falkovich2002}, these events correspond to the formation of
singularities or \textit{caustics} in the particle velocity field
\citep{Wilkinson2006,Croor2015}. Although clustering and caustics have been tied to the
centrifugal ejection of heavy particles out of vortices, they also occur in
smooth random flows that are devoid of structure
\citep{Bec2005,Wilkinson2007,sticky,Gustavsson2013,Gustavsson2014}. Consequently, the
presence of these effects does not necessarily imply that collisions sense the
structures of turbulence.

To unambiguously determine the influence of the local flow field, we must begin
with tracers which remain uniformly distributed in space. To allow for
collisions, the radii of these particles are kept (artificially) finite, while
their inertia is ignored. According to the Saffman-Turner
theory~\cite{saffman1956}, collisions should occur uniformly between any two
regions that possess the same velocity gradient magnitude, regardless of
whether these regions are vortical or straining. We now examine this
hypothesis, bearing in mind that a discrepancy will implicate flow structures
that are intrinsically more effective at causing collisions.

\begin{figure}
\includegraphics[width=.72\textwidth]{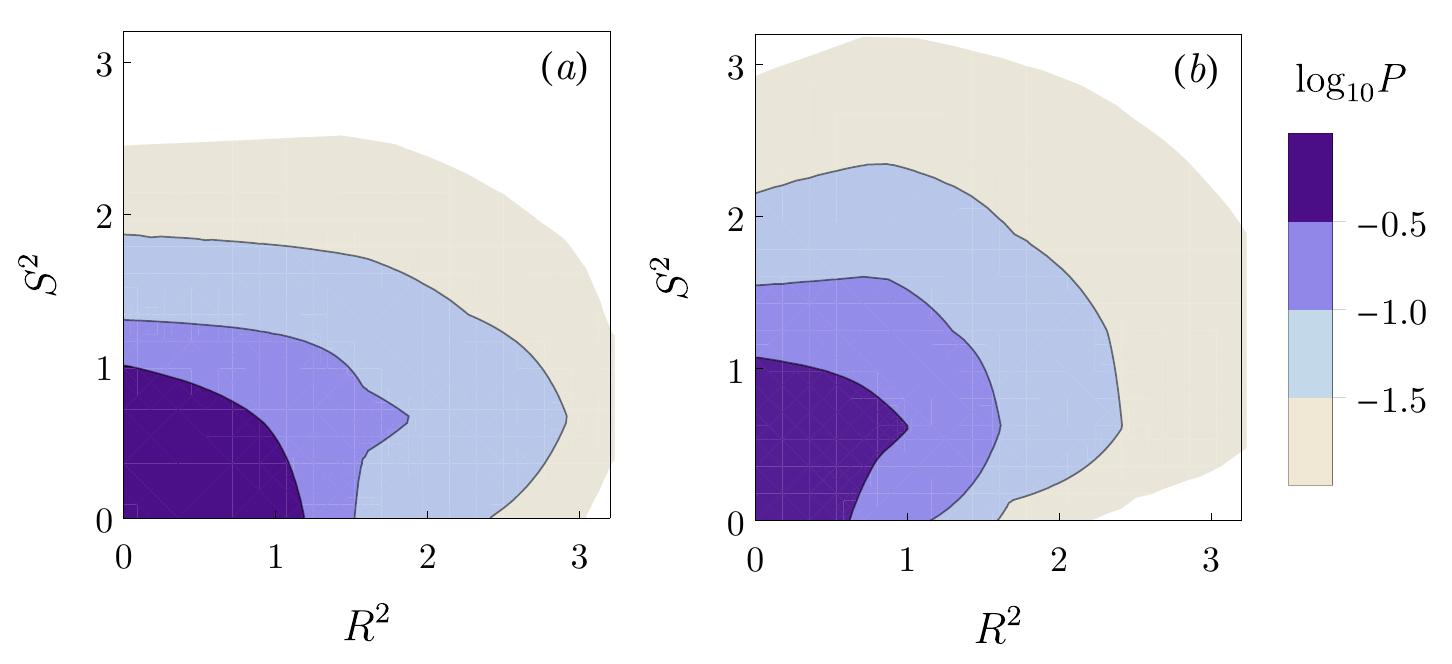}
	\caption{Bi-probability distribution functions $P(R^2,S^2)$ for inertia-less (tracer) particles, corresponding to the values of $R^2$ and $S^2$ sampled by (a) 
	particles and (b) collision locations, which show the disproportionate bias towards collisions in strain-dominated 
	regions. }
\label{fig:RS}
\end{figure}

\begin{figure*}
\includegraphics[width=0.94\textwidth]{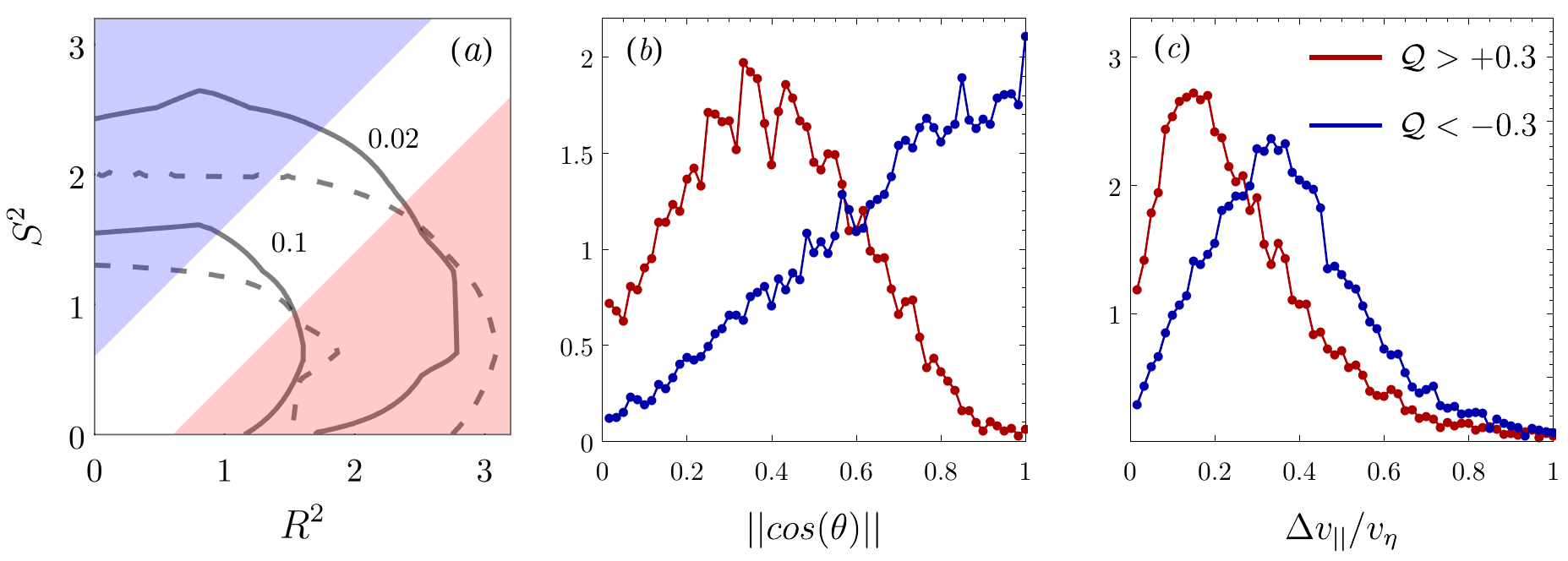}
	\caption{(a) Contours corresponding to $P(R^2,S^2) = 0.02, \;0.1$, sampled by inertia-less particles (dashed) and their collisions (solid). The blue and red shaded portions correspond to ${\mathcal Q} < -0.3$ and ${\mathcal Q} > +0.3$, and indicate regions dominated by strain and vorticity, respectively. Conditional probability distributions of  (b) the cosine of the angle of collision $\theta$, 
for data corresponding to the shaded portions of panel (a),  and (c) the approach velocity at collision $\Delta v_{||}$, normalized by the Kolmogorov velocity $v_{\eta}$. These panels clearly illustrate why straining regions are more 
effective at generating collisions. }
\label{fig:Ang}
\end{figure*}

Towards this end, we calculate the values of $R^2$ and $S^2$ along particle
trajectories, as well as at collision locations. The results for inertia-less
particles are presented as joint probability distributions functions
$P(R^2,S^2)$ in Fig.~\ref{fig:RS}a and Fig.~\ref{fig:RS}b, respectively. It is
immediately clear that collisions under-sample vortical regions ($R^2 > S^2$)
and over-sample straining regions ($S^2 > R^2$), relative to where particles reside. This is also seen in
Fig.~\ref{fig:Ang}a, which overlays contours of $P(R^2,S^2) = 0.1$ and 0.02,
for particles (dashed) and collisions (solid). The strain (vorticity) dominated
portion of this plot is shaded in blue (red), and corresponds to ${\mathcal Q}
< -0.3$ (${\mathcal Q} > +0.3$).

This surprising result is an outcome of the distinct flow topologies of these
regions which cause particles to approach each other differently.
Fig.~\ref{fig:Ang}b presents the distribution of the cosine of the collision
angle ($\theta$), for straining (blue) and vortical (red) regions. $\theta$ is
defined as the angle between the relative velocity vector ($V_{p1} - V_{p2}$)
and the separation vector ($X_{p1} - X_{p2}$) at collision. Particles in
straining regions tend to collide in a head-on or rear-end manner ($\theta
\approx 0$ or $\pi$). In either case, a large fraction of the velocity
difference between particles contributes to their rate of approach or collision
velocity ($\Delta v_{||}$). On the other hand, particles in vortices undergo
collisions that are closer to being side-on, in which case the separation
vector is nearly perpendicular to the relative velocity. This results in lower
approach velocities in vortical regions, as shown in Fig.~\ref{fig:Ang}c. Consequently, over a given time interval, fewer particles will collide in vortical regions compared to straining regions with the same magnitude of the velocity gradient and particle number density.

\begin{figure*}
\includegraphics[width=\textwidth]{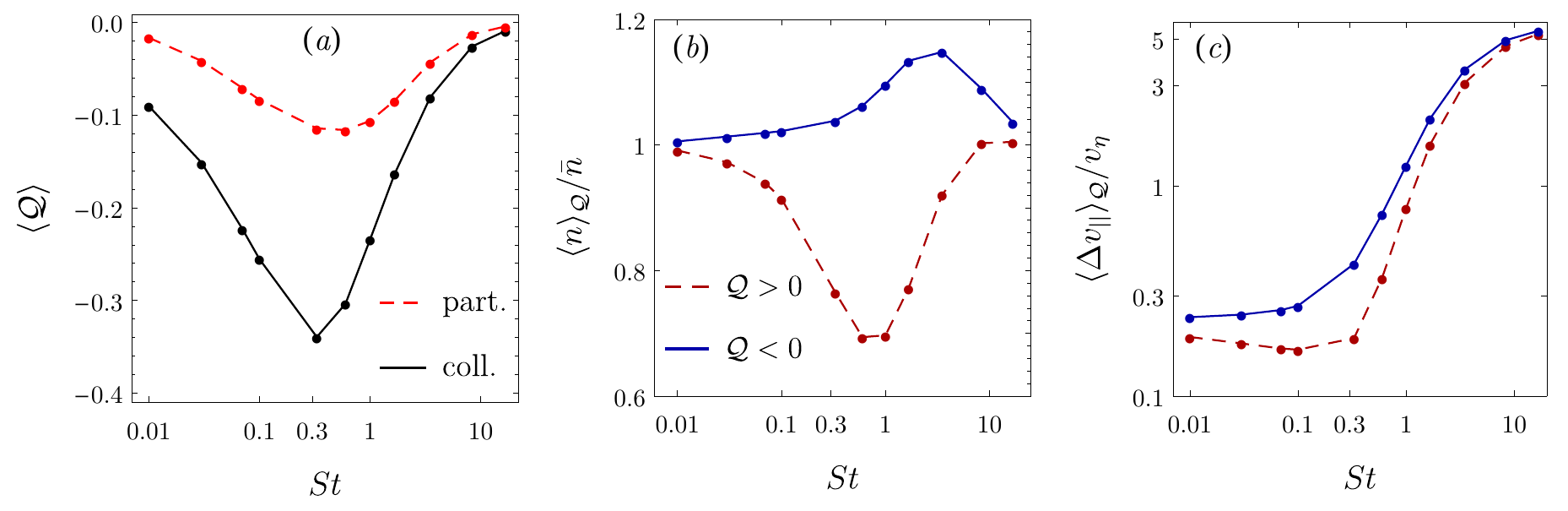}
	\caption{(a) Average ${\mathcal Q}$, sampled by particles (red-dashed) and collisions (black-solid), as a function of $\St$. (b) Average particle number density in vortical and straining regions (positive and negative $\mathcal Q$), plotted as a function of $\St$, and normalized by the domain-average number density $\bar{n}$. (c) Average approach velocity for collisions in vortical and straining regions, normalized by $v_{\eta}$}
\label{fig:den_vr}
\end{figure*}

How does particle inertia affect this picture? Figure~\ref{fig:den_vr}a presents
the average value of $\mathcal Q$ sampled by particles (dashed-red) and
their collision locations (solid-black) as a function of $\St$.  At $\St = 0$, $\langle
{\mathcal Q} \rangle$ is 0 for particles and -0.04 for collisions. Remarkably,
this offset is strongly amplified by inertia and reaches a maximum around $\St
\approx 0.3$, beyond which particles begin to de-correlate from the underlying
flow and eventually collide uniformly. The preference of \textit{inertial}
particles ($\St > 0$) to collide in straining regions has been reported
previously by Perrin and Jonker \cite{Perrin2014,Perrin2016}. 
Our results demonstrate that this effect is not fundamentally tied to particle
inertia, but rather is an amplification of a difference that exists even for
inertia-less tracers, raising the question: How does particle inertia
selectively enhance collisions in straining regions?

One possible explanation is provided by preferential concentration: Heavy
particles are centrifuged out of rotational regions, and thus tend to accumulate in straining zones just outside vortices~\citep{Croor2015}.
This causes the number density to increase in straining regions, at the expense
of vortical zones, as shown in Fig.~\ref{fig:den_vr}b. Here, $n$ is a
coarse-grained number density, obtained by dividing the domain into bins of
size $20 \eta$. The average $\mathcal Q$ in each bin is used to distinguish
between vortical ($\mathcal Q > 0$) and straining regions ($\mathcal Q < 0$)
and obtain the conditionally averaged number density. All else being equal,
higher number densities imply larger collision rates \citep{Pumir2016}.
However, we see that the maximum difference in number densities occurs near
$\St \approx 1$, which is \textit{not} where the maximum difference in $\langle
{\mathcal Q} \rangle$ is seen (Fig.~\ref{fig:den_vr}a). Hence another mechanism must
be involved.

Inertia is also known to increase the relative velocity between neighboring
particles \citep{Saw2014,James17,Bhatnagar2018}, which should result in higher collision velocities.
On examining this effect in straining and vortical regions separately, we find
that it is stronger in straining regions and, in fact, has no impact on
vortical regions for small $\St$. This is demonstrated in
Fig.~\ref{fig:den_vr}c, which presents the average values of $\Delta v_{||}$,
conditioned on $ {\mathcal Q}$. It appears that head-on (or rear-end)
collisions, which prevail in straining zones, are more amenable to being
sped-up by inertia than side-on collisions.  Notably, the maximum difference
between collision velocities occurs near $\St \approx 0.3$, which matches well
with the maximum difference in $\langle {\mathcal Q} \rangle$
(Fig.~\ref{fig:den_vr}a). Thus, larger approach rates, rather than number
densities, appear to be the primary reason for the effectiveness of straining
regions in creating collisions.

Thus far, we have considered vortical and straining regions individually.
Particle inertia, however, permits structures within a distance of $\tau_p V_p$
to influence a collision. This raises the possibility of vortical and straining
regions conspiring to enhance collisions, especially for moderately large
$\St$. For example, the geometry of vortex-strain \textit{worm-rolls} (Fig.~\ref{fig:Qstruc}) 
will cause particles in intense vortex tubes to be
rapidly ejected into strong straining sheets, where they are very likely to
collide.

We search for evidence of this effect by tracing, backward in time, all
particles that collide in straining regions. For the subset that do have
at least one particle originating from a vortical region ($Q>0$), we record (i) the time taken to
collide  after leaving the vortical region and entering the straining zone
($t_{strain}$), as indicated by $\mathcal Q$ changing sign; (ii) the strength
of the vortex, measured in terms of the maximum positive value of $\mathcal Q$
sampled inside the vortical region ($\mathcal Q_{vortex}$); (iii) the intensity of
straining at the collision location ($\mathcal Q_{col}$); and (iv) the collision
velocity. While measuring the vortex strength, we only back-track for a time of
$3 \tau_{\eta}$ within the vortical region, to ensure that the $\mathcal Q$ values
obtained are relevant to the subsequent collision. The conclusions are
insensitive to the exact value of this threshold. 
To facilitate back-tracking, we store the values of $\mathcal Q$ along all particle trajectories, at time intervals of $ \tau_{\eta}/6$, for the entire duration of the simulations. This avoids having to store the velocity field at each time step, for integrating the particle motion backward in time (as done in \cite{Wan2010}), which, given the rarity of collisions and the consequent long simulation time, would require prohibitively large amounts of computer storage.

\begin{figure*}
\includegraphics[width=\textwidth]{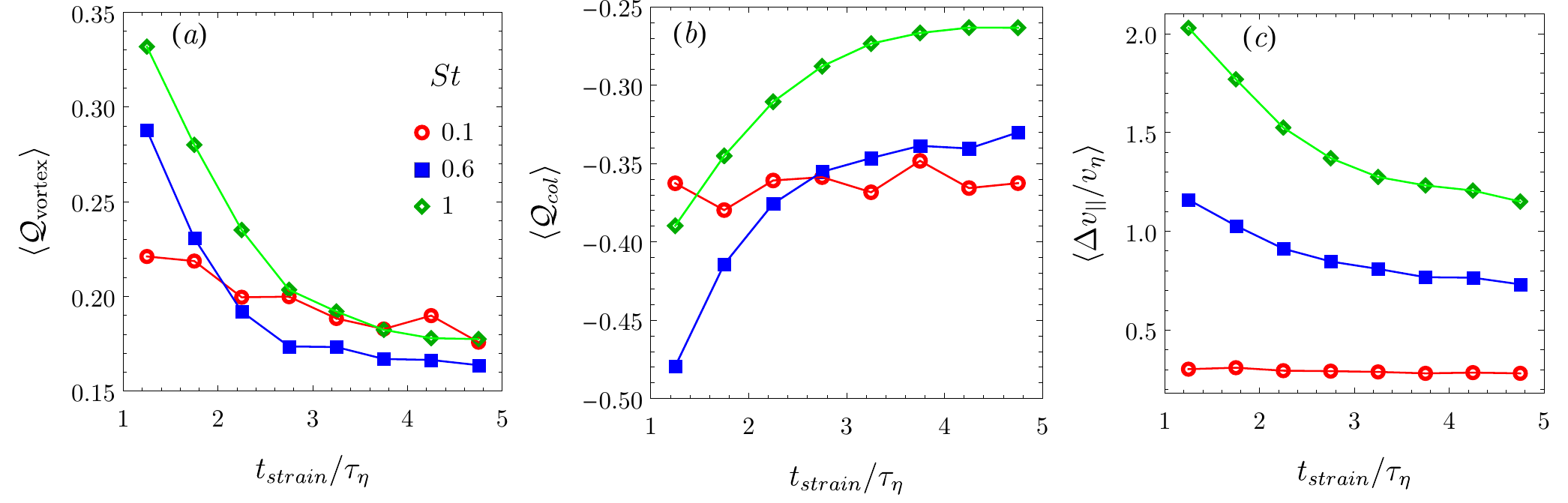}
	\caption{Representative plots of (a) the maximum value of $\mathcal Q$ sampled by the particle inside the vortical region; (b) the average value of $\mathcal Q$ at collision; and (c) the corresponding approach velocity at collisions obtained from Lagrangian tracking of particles that collide in straining regions ($\mathcal Q < 0$) after leaving a vortical region and conditionally averaged on the time taken to collide 
after entering the straining region ($t_{strain}$). These plots clearly illustrate the singular importance of vortex-strain worm-rolls as 
	discussed in the text.}
\label{fig:lag}
\end{figure*}

Figure~\ref{fig:lag} presents the results of this backward-in-time Lagrangian calculation,
conditionally averaged on the time taken to collide after leaving a vortical region, $t_{strain}$, for $\St = 1, 0.6$ and $0.1$. The data
for moderately large $\St$ (1 and 0.6), reveal the impact of vortex-strain
worm-rolls. Particles that collide quickly (small $t_{strain}$), are found to
originate from more intense vortical regions (Fig.~\ref{fig:lag}a) and to collide in
stronger straining regions (Fig.~\ref{fig:lag}b). This signature weakens
considerably for less inertial ($St = 0.1$) particles which are mildly ejected and
relax faster to the local straining flow.

The collision velocities for small $t_{strain}$ are also systematically larger
(Fig.~\ref{fig:lag}c). The standard deviation about each data point (not shown
for clarity) is of the order of the average value. Thus, several small
$t_{strain}$ collisions have very large collision velocities, indicative of
caustics/sling events, which are known to dominate the collision rate for $St >
0.5$ \citep{vobkuhle2014,Pumir2016}. Traditionally, these have been linked to
rapid ejection from vortices \citep{Falkovich2002}. Our results indicate that
this is only half the story: Straining sheets which envelope strong
vortices also contribute to generating violent collisions and enhancing
collision rates. 

All this leads one to ask how collisions are affected when structures change.
The influence of $Re_\lambda$ is particularly important to consider, as the
estimated values for natural flows are orders of magnitude larger than what can
be attained in simulations \citep{Grabowski2013,Lissauer2013}. Increasing
$Re_\lambda$ results in higher intermittency, which translates into more
intense structures (see, e.g., Ref.~\citep{burgersvortex} for a similar study
of a model stretched-vortex), but which occupy smaller volumes. These competing
effects are known to produce a non-monotonic variation of particle clustering
\citep{Onishi2014}. For collisions in particular, we have checked explicitly 
through simulation that as we increase the Reynolds numbers from $Re_\lambda=69$ to $196$, 
unsurprisingly, the differences between vortical and straining regions magnifies; therefore our 
conclusions, substantiated in this paper from simulations with $Re_\lambda=196$ hold. Nevertheless, 
we should keep in mind that $Re_\lambda=196$ is still a reasonably modest Reynolds number 
and far from those seen in atmospheric conditions. Therefore, it remains to be seen in a systematic way how this
phenomenon is affected in a higher Reynolds number flow. However, given our present understanding 
of the effect of increasing Reynolds numbers on turbulent structures, we expect that the central results 
of our work will remain unchanged and the effect that we elucidate will only show up more clearly.

Flow structures can also be significantly modified by new physical
interactions, for example,  condensation of water vapour on cloud droplets,
which releases latent heat and energizes small scales \citep{Croor2017}, and
elastic feedback from polymers that suppresses small-scale motions
\citep{Liberzon2006,Prasad2010}. Studying collisions in these complex flows is
left for future work. Furthermore, in order to clearly examine 
the effect reported in this work, we have neglected the role of gravity. We know 
that in the limit of small Froude numbers (ratio of turbulent to gravitational acceleration)~\cite{settling}, heavy droplets settle in a way 
where the fluid structures are sampled differently from the case when Froude numbers are large. 
In such low Froude number flows, it remains to be seen how the mechanism described in 
this paper is modified.

Before we conclude, it is essential to place our work in the context of
turbulent transport problems---a canonical example being that of
rain-initiation in warm clouds---which have application across the areas of
non-equilibrium statistical physics, geophysics, oceanography, astrophysics and
atmospheric sciences.  Understanding these problems demands not only an appreciation of how fast droplets sediment, collide and coalesce (in which tremendous
advances have been made in recent years) but also knowledge of
\textit{where} such processes are most likely to occur. This paper, therefore,  
contributes to a fuller understanding of this question.

\begin{acknowledgments} 
	We thank J\'er\'emie Bec, S. Ravichandran and
	Siddhartha Mukherjee for useful suggestions and discussions, which were
	facilitated in part by a program organized at ICTS: \textit{Turbulence
	from Angstroms to Light Years} (ICTS/Prog-taly2018/01).  The
simulations were performed on the ICTS clusters {\it Mowgli} and {\it
Mario} as well as
the work stations from the project ECR/2015/000361: {\it Goopy} and
{\it Bagha}.  SSR acknowledges DST (India) project
ECR/2015/000361 for financial support. 
\end{acknowledgments}

\bibliography{ref_turb_part}

\end{document}